\documentstyle[preprint,prb,aps]{revtex}
\begin{document}
\hyphenation{smaller near studied cal-cu-lations}
\hyphenation{Schweigert super-conductor super-conductors}
\hyphenation{magnetic anti-dot}
\tighten
\draft
\preprint{Accepted for publication in Phys. Rev. B}

\title{Dimensional crossover in a mesoscopic
\\ superconducting loop of finite width}
\author{V.~Bruyndoncx$^{*}$, L.~Van~Look,
M.~Verschuere, and V.~V.~Moshchalkov}
\address{Laboratorium voor Vaste-Stoffysica en Magnetisme, \\
Katholieke Universiteit Leuven, Celestijnenlaan 200 D,
B-3001 Leuven, Belgium}
\date{\today}
\maketitle

\begin{abstract}
Superconducting structures with the size of the
order of the superconducting coherence length
$\xi(T)$ have a critical temperature $T_{c}$,
oscillating as a function of the applied
perpendicular magnetic field $H$ (or flux $\Phi$).
For a thin-wire superconducting loop, the
oscillations in $T_{c}$ are perfectly periodic with
$H$ (this is the well-known Little-Parks effect),
while for a singly connected superconducting disk
the oscillations are pseudoperiodic, i.e. the
magnetic period decreases as $H$ grows. In the
present paper, we study the intermediate case: a
loop made of thick wires. By increasing the size of
the opening in the middle, the disk-like behaviour
of $T_{c}(H)$ with a quasi-linear background
(characteristic for '3-dimensional' (3D) behaviour)
is shown to evolve into a parabolic $T_{c}(H)$
background ('2D'), superimposed with perfectly
periodic oscillations. The calculations are
performed using the linearized Ginzburg-Landau
theory, with the proper normal/vacuum boundary
conditions at both the internal and the external
interface. Above a certain crossover magnetic flux
$\Phi$, $T_{c}(\Phi)$ of the loops becomes
quasi-linear, and the flux period matches with the
case of the filled disk. This dimensional transition
is similar to the 2D-3D transition for thin films in
a parallel magnetic field, where vortices enter the
material as soon as the film thickness $t > 1.8 \,
\xi(T)$. For the loops studied here, the crossover
point appears for $w\approx 1.8\,\xi(T)$ as well,
with $w$ the width of the wires forming the loop. In
the 3D regime, a "giant vortex state" establishes,
where superconductivity is concentrated near the
sample's outer interface. The vortex is then
localized inside the loop's opening.
\end{abstract}
\pacs{74.60.Ec, 74.25.Dw, 73.23.-b, 74.20.De,
74.76.-w}

\section{Introduction}
The nucleation of superconductivity in mesoscopic
samples has received a renewed interest after the
development of nanofabrication techniques, like
electron beam lithography. A superconductor is
called mesoscopic when the sample size is comparable
to the magnetic penetration depth $\lambda(T)$ and
the superconducting coherence length $\xi(T)$. In
the framework of the Ginzburg-Landau (GL) theory,
$\xi(T)$ sets the length scale for spatial
variations of the modulus of the superconducting
order parameter $|\Psi|$. The pioneering work on
mesoscopic superconductors was carried out already
in 1962 by Little and Parks~\cite{Litt62}, who
measured the shift of the critical temperature
$T_{c}(H)$ of a (multiply connected) thin-walled Sn
microcylinder (a thin-wire "loop") in an axial
magnetic field $H$. The $T_{c}(H)$ phase boundary
showed a periodic component, with the magnetic
period corresponding to the penetration of a
superconducting flux quantum $\Phi_0=h/2e$. The
Little-Parks oscillations in $T_{c}(H)$ are a
straightforward consequence of the fluxoid
quantization constraint, which was predicted by
F.~London~\cite{London50}. This condition can be
easily understood by integrating the second
GL~equation
\begin{equation}
\vec{J}=|\Psi|^2 \; \left( \vec{\nabla} \delta -
\frac{2 \pi}{\Phi_0} \vec{A} \right)= |\Psi|^2 \;\vec{v_s}
\label{GL2}
\end{equation}
along a closed contour. $J=j/(2 e \hbar/m^{\star})$
is here the normalized current density, $v_s$ is the
(normalized) superfluid velocity, and $\delta$ is
the phase of the complex order parameter
$\Psi=|\Psi| e^{\imath
\delta}$. This yields:
\begin{equation}
\oint \frac{\vec{J}}{|\Psi|^2} \cdot \vec{dl} +
2 \pi \frac{\Phi}{\Phi_0} =
\oint \vec{\nabla} \delta \cdot \vec{dl} = L \: 2 \pi
\; \; \; \; \; \; \; \; \; \; \; \; \; \; \; \;
L=\cdots \: , -2,-1,0,1,2, \: \cdots
\label{fluxoid}
\end{equation}
where we used Stokes theorem $\oint \vec{A} \cdot
\vec{dl}=\Phi$, with $\Phi$ the magnetic flux threading
the area inside the contour. In other words, when a
non-integer magnetic flux $\Phi/\Phi_0$ is applied,
a supercurrent $J$ has to be generated in order to
fulfill Eq.~(\ref{fluxoid}). The integer number $L$
is the phase winding or fluxoid quantum number and
gives the number $L$ of flux quanta $\Phi_0$
penetrating the sample. Since, for a cylindrical
geometry, the different $L$ states are
eigenfunctions of the angular (or orbital) momentum
operator as well~\cite{Daumens98}, $L$ is often
called the angular momentum quantum number. The
fluxoid quantization constraint
(Eq.~(\ref{fluxoid})) is the equivalent of the
Bohr-Sommerfeld quantum condition for Cooper
pairs~\cite{TinkBook}.

A few years later, Saint-James calculated the
$T_{c}(H)$ (or the nucleation magnetic field) of a
singly-connected cylinder~\cite{SJ65cyl} (a
mesoscopic "disk"). Taking into account the analogy
with the situation of a semi-infinite
superconducting slab in contact with
vacuum~\cite{SJ63lin}, the critical field was called
$H_{c3}(T)$ in this case, since superconductivity
nucleates initially {\em near the sample interface}.
As we will show further in this paper, a substantial
enhancement of the nucleation field, above the
$H_{c3}(T)$ for a semi-infinite slab, can be
obtained in mesoscopic samples, where the
surface-to-volume ratio is large. In such case,
surface effects play the role of the bulk effects.
Therefore, we will rather use the notation
$H_{c3}^{*}(T)$. For a disk, the limiting value
$H_{c3}^{*}(T)=H_{c3}(T)=1.69\; H_{c2}(T)$ is found
as $\Phi \rightarrow \infty$ (or the radius of the
disk $\rightarrow \infty$). The field $H_{c2}$ is
the upper critical field of a bulk Type~II
superconductor, at which the Abrikosov vortex
lattice is formed.

The $T_{c}(H)$ phase boundary (or $H_{c3}^{*}(T)$)
of the disk shows, just like for the usual
Little-Parks~effect in a multiply connected sample
(loop), an oscillatory behaviour. In this case, the
oscillation period of $T_{c}(H)$ is not constant,
but slightly decreases as $H$ increases. When moving
along $T_{c}(H)$, superconductivity concentrates
more and more near the sample interface as $H$
grows. A giant vortex state is formed: a "normal"
core carries $L$ flux quanta, and the 'effective'
loop radius increases, resulting in a decrease of
the magnetic oscillation period. Here as well,
fluxoid quantization (Eq.~(\ref{fluxoid})) is
responsible for the oscillations of $T_{c}$ versus
$H$. An experimental verification of these
predictions was carried out later on by Buisson {\it
et al.}~\cite{Buisson90} and by Moshchalkov {\it et
al.}~\cite{VVM95Nature}.

Currently, many different sample topologies are
studied: superconducting
networks~\cite{Pannetier91}, antidot
systems~\cite{Bezr95PRBonehole,Bezr9495,Moshchalkov96},
samples consisting of sharp
corners~\cite{Corners,SchweigertPRB98}, etc. With
the use of sub-micron Hall probe
microscopy~\cite{GeimAPL} for measuring the magnetic
response of a superconductor, it has become possible
to probe samples deep in the superconducting
state~\cite{GeimNatures} (i.e. at temperatures far
below $T_{c}$). A lot of recent theoretical
activities have been devoted to the magnetization
and vortex configurations in superconducting disks
of different
sizes~\cite{SchweigertPRB98,magntheory}.

Loop structures have also been studied extensively
in the past years. A large portion of the
theoretical research has been focused on the
transition between two different fluxoid states $L
\rightarrow L+1 \,$~\cite{Bezr95PRBonehole,BergerV},
but also experiments, among which susceptibility
measurements close to $T_{c}$~\cite{Zhang97},
studies of energy gap spectrum using a
single-electron transistor~\cite{Sato}, and scanning
Hall probe measurements on ensembles of mesoscopic
Al loops~\cite{Davidovic}, were carried out.

As it has already been mentioned, the two limiting
cases, "thin-wire loop" and "disk", are well
understood as far as the nucleation field
$H_{c3}^{*}$ is concerned, but the intermediate case
is not. In the early paper from Saint-James and
de~Gennes~\cite{SJ63lin}, $H_{c3}^{*}(T)$ has been
calculated also for a {\em film exposed to a
parallel magnetic field}, where surface
superconductivity can grow along two
superconductor/vacuum interfaces. For low magnetic
fields, the two surface superconducting sheaths
overlap, and, as a result, $T_{c}$ versus $H$
becomes parabolic, which is characteristic for 2D
behaviour. When increasing the field, a crossover to
a linear $T_{c}(H)$ dependence (3D) occurs at $t
\approx 2 \, \xi(T)$, with $t$ the film thickness.
Shortly after, it was shown that vortices start to
nucleate in the film at this dimensional crossover
point ($t = 1.8 \,
\xi(T)$)~\cite{FinkSchultenscrossover}.

The goal of the present paper is to study the phase
boundary $T_{c}(H)$ of loops made of finite width
wires (or disks with an opening in the middle,
increasing in size). In a Type-II material,
superconductivity is expected to be enhanced, with
respect to the bulk ($H_{c3}^{*}(T)>H_{c2}(T)$),
both at the external and the internal sample
surfaces. As for a film in a parallel field, a 2D-3D
dimensional crossover can be anticipated, since the
loops may be simply considered as a film, which is
bent such that its ends are joined together. We
calculate the phase boundary $T_{c}(H)$ as the
ground state solution of the linearized first
GL~equation with two superconductor/vacuum
interfaces. This calculation has been suggested
already by several
authors~\cite{Bezr9495,Benoist97}, but to the best
of our knowledge, it has not been carried out so
far. We will show that the $T_{c}(\Phi)$ of the
loops, for low applied magnetic flux (corresponding
to the 2D regime), can be described within a simple
London picture, where the modulus of the order
parameter $|\Psi|$ is spatially constant. As the
flux increases, the background depression of
$T_{c}(\Phi)$ is changed from parabolic (2D) to
quasi-linear (3D), which indicates the formation of
a giant vortex state, where only a surface sheath
close to the sample's outer interface is in the
superconducting state. Moreover, the oscillation
period of $T_{c}(\Phi)$ becomes identical for the
loops as for the full disk, as soon as the
transition to 3D behaviour has taken place.

\section{The linearized GL equation}
The linearized first GL equation to be solved in
order to find $T_{c}(H)$ is:
\begin{equation}
\frac{1}{2m^{\star}}(-i \hbar\vec{\nabla}-2 e
\vec{A})^{2}\Psi = |-\alpha| \, \Psi \; .
\label{glfree1}
\end{equation}
This equation is formally identical to the
Schr\"{o}dinger equation for a particle with a
charge $2 e$ in a magnetic field. Here, at the onset
of superconductivity $T \approx T_{c}(H)$, the
nonlinear term $\beta |\Psi|^{2} \, \Psi \ll
|-\alpha| \, \Psi$ can been omitted, and
demagnetization effects along the field direction do
not need to be considered. In this regime, the
$z$-dependence disappears from the equations and
therefore an infinitely long cylinder and a disk
have identical $T_{c}(H)$ boundaries. It is further
assumed that the penetration depth $\lambda(T) \gg$
sample size, so that $\mu_0
\vec{H}=rot\vec{A}$. It is important to note that
$H$ is the {\em applied magnetic field}, so fields
induced by supercurrents $J$ (Eq.~(\ref{fluxoid}))
are not taken into account here. The more Type~II
(the higher $\kappa$) the superconducting material
is, the larger range of validity our calculations
have. The eigenenergies $|-\alpha|$ can be written
as:
\begin{equation}
|- \alpha|=\frac{\hbar^{2}}{2 m^{\star} \:
\xi^{2}(T)}=\frac{\hbar^2}{2 m^{\star} \: \xi^{2}(0)}
\left( 1- \frac{T}{T_{c0}} \right) \; ,
\label{glalpha}
\end{equation}
$T_{c0}$ being the critical temperature in zero
magnetic field. From the energy eigenvalues of
Eq.~(\ref{glfree1}), the lowest Landau level
$|-\alpha_{LLL}(H)|$ is directly related to the
highest possible temperature $T_{c}(H)$, for which
superconductivity can exist.

By varying the topology of the sample
("nanostructuring"), the lowest Landau level
$|-\alpha_{LLL}|$ can tuned by confinement of the
superconducting condensate. Several examples of this
concept can be found in
Refs.~\cite{Pannetier91,HanNano}. Indeed, the
solution $\Psi$ of Eq.~(\ref{glfree1}) has to
fulfill the Neumann boundary condition
\begin{equation}
\left. (- \imath \hbar \vec{\nabla} - 2 e \vec{A})
\Psi \right|_{\perp , b}=0 \; ,
\label{glbound}
\end{equation}
at the sample interfaces $b$. This requirement
guarantees that the supercurrent does not have a
component perpendicular to a superconductor/vacuum
interface.

For the loop geometries, we choose the cylindrical
coordinate system $(r,\varphi)$ and the gauge
$\vec{A}=(\mu_0 H r/2)\,
\vec{e}_{\varphi}$, where $\vec{e}_{\varphi}$ is the
tangential unit vector. The exact solution of the
Hamiltonian (Eq.~(\ref{glfree1})) in cylindrical
coordinates takes the following
form\cite{Bezr9495,Benoist97,Dingle,VVMQiu}:
\begin{eqnarray}
\Psi(\Phi,\varphi)& = & e^{- \imath L \varphi}
\left( \frac{\Phi}{\Phi_0} \right)^{L/2}
\exp \left( - \frac{\Phi}{2 \, \Phi_0}
\right) K(-n,L+1,\Phi / \Phi_0) \label{psikummer} \\
K(a,c,y)& = & c_1 \, M(a,c,y)+ c_2 \, U(a,c,y) \, .
\nonumber
\end{eqnarray}
Here $\Phi=\mu_0 H \pi r^2$ is the applied magnetic
flux through a circle of radius $r$. The number $n$
determines the energy eigenvalue. Most generally,
the function $K(a,c,y)$ can be any linear
combination of the two confluent hypergeometric
functions (or Kummer functions) $M(a,c,y)$ and
$U(a,c,y)$~\cite{Abramowitz}, but the sample
topology puts a constraint on $c_1, \, c_2$, and
$n$, via the boundary condition
(Eq.~(\ref{glbound})).

The eigenenergies of Eq.~(\ref{glfree1}) (the Landau
levels) are~\cite{VVMQiu}:
\begin{equation}
|-\alpha| = \frac{2 e \hbar \, \mu_0 H}{2
\, m^{\star}}
\left( 2 n +1 \right)=\hbar \, \omega
\left( n+\frac{1}{2} \right) \, .
\label{eperpend}
\end{equation}
where $\omega=2 e \mu_0 H / m^{\star}$ is the
cyclotron frequency. {\em The parameter $n$ depends
on $L$ and is not necessarily an integer number}, as
we shall see later. With Eq.~(\ref{glalpha}) this
can be rewritten as:
\begin{equation}
\frac{r_o^2}{\xi^2 (T_{c})} =
\frac{r_o^2}{\xi^2(0)} \,
\left( 1- \frac{T_{c}(H)}{T_{c0}} \right)
= 4 \, \left( n+ \frac{1}{2}
\right) \, \frac{\Phi}{\Phi_0}
= \epsilon(H_{c3}^{*}) \, \frac{\Phi}{\Phi_0}  \, ,
\label{hcscaled}
\end{equation}
where $\Phi=\mu_0 H \pi r_o^2$ is arbitrarily
defined.

The bulk Landau levels can be found when
substituting $n=0,1,2,\: \cdots$~ in
Eqs.~(\ref{eperpend}) and~(\ref{hcscaled}), meaning
that the lowest level $n=0$ corresponds to the upper
critical field $\mu_0 H_{c2}(T)=\Phi_0/\left( 2 \pi
\xi^2(T)\right)$. Let us note that the lowest
Landau level ($n=0$) for a bulk superconductor is
degenerate in the phase winding number $L$, and
therefore the eigenfunction can be expanded as
$\Psi=\sum c_L
\Psi_L$. Interference patterns between the different
functions $\Psi_L$ give rise to aperiodic vortex
states~\cite{MoshDhal93}.

The boundary condition (Eq.~(\ref{glbound})), in
cylindrical coordinates, can be simply written as:
\begin{equation}
\left. \frac{\partial |\Psi(r)|}{\partial r}
\right|_{r=r_o} \, ,
\label{boundscal}
\end{equation}
with a superconductor/vacuum interface at a radius
$r_o$. Using $\frac{d M(a,c,y)}{d y}=
\frac{a}{c} M(a+1,c+1,y)$ and
$\frac{d U(a,c,y)}{d y}= -a \, U(a+1,c+1,y)$ for the
derivatives of the first and second type of Kummer
functions, respectively~\cite{Abramowitz}, and
inserting Eq.~(\ref{psikummer}) into
Eq.~(\ref{boundscal}) gives:
\begin{eqnarray}
c_1 \, \left[ \left( L - \frac{\Phi}{\Phi_0} \right)
\, M \left( -n,L+1,\Phi/\Phi_0 \right) -
\frac{2 \, n}{L+1} \,\frac{\Phi}{\Phi_0} \,
M \left( -n+1,L+2,\Phi/\Phi_0\right) \right] +
\label{boundloop} \\
\left. c_2 \, \left[ \left( L -
\frac{\Phi}{\Phi_0}
\right)
\, U \left( -n,L+1,\Phi/\Phi_0 \right) +
2 \, n \,\frac{\Phi}{\Phi_0} \, U \left(
-n+1,L+2,\Phi/\Phi_0\right) \right] \, \,
\right|_b=0\, ,\nonumber
\end{eqnarray}
which has to be solved numerically for each integer
value of $L$, resulting in a set of values
$n(L,\Phi)$, with $\Phi=\mu_0 H \pi r_o^2$.

For a disk geometry~\cite{SJ65cyl,Buisson90,VVMQiu},
we have to take $c_2=0$ in Eqs.~(\ref{psikummer})
and (\ref{boundloop}) in order to avoid the
divergency of $U(a,c,y \rightarrow 0)=\infty$ at the
origin. Selecting the lowest Landau level at each
value $\Phi$, one ends up with a cusp-like
$T_{c}(H)$ phase boundary~\cite{SJ65cyl}, which is
composed of values $n<0$ in Eq.~(\ref{eperpend}),
thus leading to $H_{c3}^{*}(T)>H_{c2}(T)$. A similar
calculation was performed for a single circular
microhole in a plane film
("antidot")~\cite{Bezr9495}, where $c_1=0$ in
Eqs.~(\ref{psikummer}) and (\ref{boundloop}), since
$M(a,c,y \rightarrow \infty)=\infty$. Here as well,
the lowest Landau level consists of solutions with
$n<0$. At each cusp in $T_{c}(\Phi)$, the system
makes a transition $L
\rightarrow L \pm 1$, i.e. a vortex enters or is
removed from the sample.

The loops we are currently studying have two
superconducting/vacuum interfaces, one at the outer
radius $r_o$, and one at the inner radius $r_i$.
Consequently, the boundary condition
(Eq.~(\ref{boundloop})) has to be fulfilled at both
$r_o$, and $r_i$. As a result, we have a system of
two equations and two variables $n$ and $c_2$
($c_1=1$ is chosen), which we solved for different
values of $x=r_i/r_o$. {\em In the rest of the paper
we will define $\Phi=\mu_0 H \pi r_o^2$, where $r_o$
always means the outer loop radius.}

\section{The London limit}
The usual description of the
Little-Parks~effect~\cite{TinkBook,dGABook} is given
in terms of the London limit, where $|\Psi|$ is
spatially constant. This approximation, of course,
is valid when the wire, forming the loop, is very
thin ($x\approx 1$), or when we define the loop
aspect ratio $z$ as
\begin{equation}
z=\frac{r_o-r_i}{r_o+r_i}=\frac{1-x}{1+x} \, ,
\end{equation}
this condition implies $z \ll 1$.

The solution of the linearized GL~equation
(Eq.~(\ref{glfree1})) becomes~\cite{Groff64}:
\begin{equation}
\frac{r_m^2}{\xi^2 (T_{c})} =
\frac{r_m^2}{\xi^2(0)} \,
\left( 1- \frac{T_{c}(H)}{T_{c0}} \right)
= \left( \frac{\Phi_m}{\Phi_0} \right)^2 \,
\left( 1+z^2 \right) - 2 \, L \,
\frac{\Phi_m}{\Phi_0}+ \frac{L^2}{2 \, z} \ln
\left( \frac{1+z}{1-z} \right) \, ,
\label{grofftc}
\end{equation}
with $\Phi_m= \mu_0 H \pi r_m^2$, $r_m$ is the mean
radius of the loop. Note that this definition of
flux $\Phi_m$ is different from $\Phi$ in the
previous section. The lowest eigenvalues are
obtained when $L$ is the integer number closest to
$-2 \, (\Phi_m/\Phi_0) \, z / \ln x$. We will
further compare this equation with our more exact
results, calculated from the scheme presented in the
previous section.

Since in the original paper by Groff and
Parks~\cite{Groff64}, thin-wire loops ($z \ll 1$)
were investigated, the logarithm was expanded in a
Taylor series, which gives, up to the order $z^2$:
\begin{equation}
\frac{r_m^2}{\xi^2 (T_{c})} =
\frac{r_m^2}{\xi^2(0)} \,
\left( 1- \frac{T_{c}(H)}{T_{c0}} \right)
= \left( L - \frac{\Phi_m}{\Phi_0} \right)^2 +
\frac{4}{3} \, z^2 \,
\left( \frac{\Phi_m}{\Phi_0} \right)^2 \, .
\label{groffexp}
\end{equation}
The first term on the right hand side of
Eq.~(\ref{groffexp}) is the periodic part of the
$T_{c}$ reduction (i.e. the Little-Parks effect),
while the second term is a monotonic parabolic
background, which is identical to the $T_{c}(H)$
expression for a plane film of thickness $t=2 \, z
\, r_m$ in a parallel magnetic
field~\cite{TinkBook}. In Ref.~\cite{Groff64}, a
substitution is performed, which splits the right
hand side of Eq.~(\ref{grofftc}) in a periodic term
and a monotonic background $T_{c}$. The latter
becomes:
\begin{equation}
\frac{r_m^2}{\xi^2 (T_{c})} =
(1+z^2) \, \left\{ \left( \frac{1+z^2}{2 \, z}
\right) \, \ln \left( \frac{1+z}{1-z} \right) -1
\right\}
\, \left( \frac{\Phi_m}{\Phi_0} \right)^2 \, ,
\label{GPback}
\end{equation}
which is parabolic with $\Phi_m$. This substitution
is only valid for thin-wire loops, and
Eq.~(\ref{GPback}) transforms of course into the
last term of Eq.~(\ref{groffexp}) for $z\ll 1$.

\section{Results}
Fig.~\ref{elevels} shows the Landau level scheme
(dashed lines) calculated from
Eqs.~(\ref{hcscaled}), and~(\ref{boundloop}) for
loops with a different inner radius $x=r_i/r_o$
(a)~$x=0.1$, (b)~$x=0.3$, (c)~$x=0.5$, (d)~$x=0.7$,
and (e)~$x=0.9$. The applied magnetic flux $\Phi=
\mu_0 H \pi r_o^2$ is defined with respect to the outer
sample area. The $T_{c}(H)$ boundary (or
$|-\alpha_{LLL}(H)|$) is composed of $\Psi$
solutions with a different phase winding number $L$
and is drawn as a solid cusp-like line in
Fig.~\ref{elevels}. At $\Phi
\approx 0$, the state with $L=0$ is formed at $T_{c}(H)$
and one by one, consecutive flux quanta $L$ enter
the loop as the magnetic field increases. Each state
$L$ approximately has a parabolic dependence
$|-\alpha(H)|$, close to $T_{c}(H)$. Like for the
disk~\cite{GeimNatures,Benoist97}, where $\Phi
\approx \Phi_0 \, (L+L^{1/2})$, we have here as
well $L \, \Phi_0 \leq \Phi$, indicating the overall
diamagnetic response of the sample. As $x$
increases, the oscillations in $T_{c}(\Phi)$ change
from cusp-like, to very pronounced local extrema for
$x=0.9$. In the limit of vanishing wire width ($x
\rightarrow 1$), $L$ is the integer closest to
$\Phi_m/\Phi_0$, and therefore the response of the
loop is alternating between diamagnetic and
paramagnetic, as the flux varies.

The solid and dotted straight lines in
Fig.~\ref{elevels} are the bulk upper critical field
$H_{c2}(T)$ and the surface critical field
$H_{c3}(T)$ for a semi-infinite slab, respectively.
In these units the slopes of the curves (see
Eq.~(\ref{hcscaled})) are $\epsilon=2$ for $H_{c2}$
(substitute $n=0$ in Eq.~(\ref{hcscaled})) and
$\epsilon=2/1.69$ for $H_{c3}$. The ratio
$\eta=\epsilon(H_{c2})/\epsilon(H_{c3})=1.69$
corresponds then to the enhancement factor
$H_{c3}(T)/H_{c2}(T)$ at a constant temperature. For
the loops we are studying here
$\eta=\epsilon(H_{c2})/\epsilon(H_{c3}^{*})$ is no
longer a constant, but varies with the magnetic
field.

The energy levels below the $H_{c2}$ line (solid
straight line in Fig.~\ref{elevels}) could be found
by fixing a certain $L$, and solving
Eq.~(\ref{boundloop}) for a small $\Phi$, until a
set $(n,c_2)$ is found with $n<0$. These values were
always put in as starting values for a slightly
higher $\Phi$. A trivial solution of
Eq.~(\ref{boundloop}) is obtained for
$n=0,1,2,\cdots \, $. Both confluent hypergeometric
functions reduce to $M(-N,L+1,\Phi/\Phi_0)=1$ and
$U(-N,L+1,\Phi/\Phi_0)=1$, and thus $c_2=-c_1$.
Inserting this into Eq.~(\ref{psikummer}) gives
$\Psi(\Phi,\varphi)=0$ everywhere. With this method,
we were able to find solutions with $n<0$
numerically. Note that the lowest Landau level
always has a lower energy $|-\alpha(\Phi)|$ than for
a semi-infinite superconducting slab, which implies
$H_{c3}^{*}(T)>H_{c3}(T)$.

The dash-dotted curve in Fig.~\ref{elevels} gives
the result obtained with the London
limit~\cite{Groff64} (see Eq.~(\ref{grofftc})). In
Fig.~\ref{elevels}a ($x=0.1$) the deviation from the
exact solution of Eq.~(\ref{glfree1}) appears
already for $L \geq 1$. For low flux, the result
from Eq.~(\ref{grofftc}), has clearly higher energy
$|-\alpha|$ than the surface critical field
$H_{c3}(T)$. At $\Phi
\approx 7 \, \Phi_0$, this curve even crosses the
bulk $H_{c2}(T)$ line, which is clearly unphysical.
For $x=0.3$ (Fig.~\ref{elevels}b) the London limit
is valid up to $\Phi \approx 4
\, \Phi_0$, up to $\Phi \approx 8 \, \Phi_0$ for $x=0.5$
(Fig.~\ref{elevels}c), and for $x=0.7$
(Fig.~\ref{elevels}d) it is a good approximation in
the whole flux interval of our calculations.
Finally, for $x=0.9$ (Fig.~\ref{elevels}e) the
assumption of a spatially constant $|\Psi|$ gives a
$T_{c}(\Phi)$, which can not be distinguished from
the exact solution of Eq.~(\ref{glfree1}).

In Fig.~\ref{elevels}a ($x=0.1$) the background
depression of $T_{c}$ is {\em quasi-linear}, just
like for the case of a filled disk, for $x=0.3$
(Fig.~\ref{elevels}b) $T_{c}(H)$ has a rather {\em
parabolic} background for flux $\Phi \lesssim 7
\, \Phi_0$, and becomes quasi-linear at higher flux.
This becomes more clear for $x=0.5$
(Fig.~\ref{elevels}c). Here, the crossover point
from parabolic to quasi-linear appears at about
$\Phi \approx 14 \, \Phi_0$. In Figs.~\ref{elevels}d
and~\ref{elevels}e finally ($x=0.7$, $x=0.9$), the
background is parabolic in the entire flux regime
and can be very accurately described by
Eq.~(\ref{GPback}). Simultaneously, as $x$
increases, the cusps in $T_{c}(H)$ become more and
more pronounced, until the usual Little-Parks~effect
is recovered for $x=0.9$ (see Eq.~(\ref{groffexp})),
where sharp local minima and clear maxima in
$T_{c}(H)$ are seen.

\section{Discussion}
In order to study the spatial variation of the order
parameter $|\Psi|$, we will use Abrikosov's
definition of the flatness parameter
$\beta_A=<|\Psi|^4>/<|\Psi|^2>^2$, where the
brackets indicate the average over the actual sample
area~\cite{Abrikosov} (not including the middle
opening). Abrikosov introduced this parameter
$\beta_A$ in order to find the most favorable order
parameter distribution, in a bulk system, near the
$H_{c2}$ line. In this language, $\beta_A=1$ means a
completely flat profile of $|\Psi|$, corresponding
to the London limit. For comparison, we mention that
$\beta_A=1.16$ for a triangular Abrikosov vortex
lattice.

In Fig.~\ref{Profilesx01} we plot the modulus of the
order parameter $|\Psi|$ for the case $x=0.1$, at
$\Phi=9 \, \Phi_0$, for the states a) $L=2$, b)
$L=3$, c) $L=4$, d) $L=5$, and e) $L=6$. The
$|\Psi|$ values have been normalized to $1$ at the
sample's outer interface $r=r_o$. The dark area is
the region outside the sample. The maximum in
$|\Psi|$ for $L=2$ lies at $r\approx (r_i+r_o)/2$,
and gradually shifts to the outer sample edge as $L
\rightarrow 6$, which is the ground state solution
of Eq.~(\ref{glfree1}). The spatial modulation of
$|\Psi|$ is considerable ($\beta_A > 1.16$) for all
$|\Psi|$ patterns shown here. At $T=T_{c}(\Phi)$
(Fig.~\ref{Profilesx01}e), ($L=6, \, \beta_A=1.59$)
the sample is in the "giant vortex state", with a
normal core containing $6$ flux quanta $\Phi_0$, and
a surface superconducting sheath at the outer sample
edge.

The $|\Psi|$ profiles for a loop with a larger inner
radius ($x=0.5$) are shown in
Fig.~\ref{Profilesx05}. Here as well, we have chosen
$\Phi = 9
\, \Phi_0$, and the same normalization
$|\Psi(r_o)|=1$. a)~$L=2$, b)~$L=3$, c)~$L=4$,
d)~$L=5$, and e)~$L=6$. For $L=2$
(Fig.~\ref{Profilesx05}a), $|\Psi|$ has a maximum at
$r=r_i$. For higher $L$ the order parameter
distribution flattens until it reaches
$\beta_A=1.06$ for $L=4$. Then, for the ground state
energy ($L=5$) (Fig.~\ref{Profilesx05}d), the
maximum in $|\Psi|$ moves outward, but we should
notice that for this state $\beta_A=1.01$ only,
which means that superconductivity nucleates in a
quasi-uniform way. Indeed, the exact $T_{c}(\Phi)$
and the London limit result are still very close to
each other at $\Phi= 9 \, \Phi_0$ (see
Fig.~\ref{elevels}c). Although multiple flux quanta
$L$ are threading the middle opening of the loop, we
can not, in the strict sense, speak about a giant
vortex state here. First of all, there is no real
'normal core' within the sample area, and secondly
we are not dealing with a surface superconducting
state in this case. The surface-to-volume ratio is
so large that the whole sample area becomes
superconducting at once. On the contrary, for the
disk, strong spatial gradients of $|\Psi|$ are
responsible for the spontaneous breaking of
superconductivity in the giant vortex core, while
only a surface sheath is superconducting.

It is worth noting that for all of the states shown
in Fig.~\ref{Profilesx01} ($x=0.1$), the width of
the wire $w=r_o-r_i=(1-x)\, r_o >2.7 \,
\xi(T)$ (see also Fig.~\ref{elevels}a), while for
the loop with $x=0.5$ the different $L$ states from
Fig.~\ref{Profilesx05} correspond to $w<2.1 \,
\xi(T)$ (see also Fig.~\ref{elevels}c). At this point
we want to remind that in a thin film of thickness
$t$ in a parallel field $H$, a dimensional crossover
is found at $t=1.84 \,
\xi(T)$. For low fields (high $\xi$) $T_{c}(H)$
is parabolic (2D), and for higher fields vortices
start penetrating the film and consequently
$T_{c}(H)$ becomes linear
(3D)~\cite{FinkSchultenscrossover}. In
Figs.~\ref{elevels}a,~b, and~c the small arrow
indicates the point on the phase diagram
$T_{c}(\Phi)$ where $w=1.84 \, \xi(T)$. For loops
with larger $x$ this point lies outside the flux
regime of the calculations. For the loops as well,
the dimensional transition shows up approximately at
this point, although the vortices are not
penetrating the sample area in the 3D regime.
Instead, the middle loop opening contains an integer
number of flux quanta $L \, \Phi_0$. In the present
case, the 2D-3D crossover roughly occurs at the
value $\Phi$ where the London limit result
(Eq.~(\ref{grofftc})) fails to describe the exact
$T_{c}(\Phi)$.

In order to compare the flux periodicity of
$T_{c}(\Phi)$, we have replotted, in
Fig.~\ref{enhancement}, the lowest energy levels of
Fig.~\ref{elevels} as
$\eta^{-1}=\epsilon(H_{c3}^{*})/\epsilon(H_{c2})$,
which is the inverse enhancement factor at a
constant temperature. In this representation, the
dotted horizontal line at $\eta^{-1}=0.59$
corresponds to the surface critical field line
$H_{c3}(T)$. The nucleation field of a disk
$H_{c3}^{*}(T)>1.69 \, H_{c2}(T)$ (i.e.
$\eta>1.69$), and for a circular microhole in an
infinite film ("antidot")~\cite{Bezr9495}
$H_{c3}^{*}(T)<1.69 \, H_{c2}(T)$ ($\eta<1.69$). As
$\Phi$ grows (the radius goes to infinity) the
$H_{c3}^{*}(T)$ of both the disk and the antidot
approaches the $H_{c3}(T)$ line.

At this point, we would like to come back to the
explanation why the notation $H_{c3}^{*}(T)$ was
used for the nucleation field. For a thin film
(thickness $t$) with the field $H$ applied parallel
to the surfaces, $T_{c}(H)$ is found from the second
term of Eq.~(\ref{groffexp}). With $t=2 \, z \, r_m$
and $\mu_0 H_{c2}=\Phi_0 / (2 \, \pi \, \xi^2(T))$,
this becomes
\begin{equation}
H_{c3}^{*}(T)=\frac{\sqrt{12} \, \,\xi(T)}{t} \,
H_{c2}(T) = 2.04 \, \frac{\xi(T)}{t} \, H_{c3}(T) \,
,
\label{filmparallel}
\end{equation}
which implies directly $H_{c3}^{*}(T)>H_{c3}(T)$ for
very thin films $t< 2.04
\, \xi(T)$~\cite{RothwarfFinkhc4}. Note that
$t=w \approx 2.04 \, \xi(T)$ is very close to the
2D-3D crossover point in films and in loops. Of
course, the possibility for nucleation fields
$H_{c3}^{*}>H_{c3}$ is not very special, but it
still creates a lot of confusion. In thin films, for
example, the critical field $H_{c3}^{*}(T)$ is often
denoted as $H_{c2,\perp}$ in a perpendicular
magnetic field, and $H_{c2,\parallel}$ in a parallel
field, which would mean
$H_{c2,\parallel}(T)>H_{c3}(T)$ for $t< 2.04 \,
\xi(T)$ (Eq.~(\ref{filmparallel})).
Van Gelder studied the $H_{c3}^{*}$ of a
semi-infinite film which is bent over a certain
angle (a "wedge") and even called it $H_{c4}$, since
it exceeds $H_{c3}$ at small angles~\cite{Gelder68}.
Since this is just a finite size effect and indeed
no new nucleation mechanism is involved, the
existence of a $H_{c4}>H_{c3}$ was called a
misinterpretation by Fink~\cite{RothwarfFinkhc4}.
For all these reasons, it is safe to use the
notation $H_{c3}^{*}$ for the nucleation field.
Enhancing the nucleation field $H_{c3}^{*}(T)$ can
be realized by confining the superconducting
condensate in a mesoscopic sample. The smaller the
sample size, the lower the Landau level
$|-\alpha_{LLL}(H)|$ becomes, when Neumann boundary
conditions (Eq.~(\ref{glbound})) are imposed. The
idea of having $H_{c3}^{*}>H_{c3}=1.69
\, H_{c2}$ was used to explain anomalously high
values of $H_{c3}^{*}$ and attribute this to the
existence of small surface irregularities (of the
order of $\xi$), impurities, grain boundaries or
concentration gradients near the
interface~\cite{LowellJoiner}.

For all the loops we study here, the presence of the
outer sample interface automatically implies that
$H_{c3}^{*}(T)>H_{c3}(T)$ is enhanced ($\eta>1.69$),
with respect to the case of a flat
superconductor-vacuum interface. For loops with a
small $x$, the $T_{c}(\Phi)$ boundary very rapidly
collapses with the $T_{c}(\Phi)$ of the dot (for
$\Phi > 4 \,
\Phi_0$) ($\eta$ becomes the same). Since both the
flux periodicity and the background depression of
$T_{c}$ become identical to the disk, a giant vortex
state can be anticipated for the loop as well. The
presence of the opening in the sample is not
relevant for the giant vortex formation in the high
flux regime. Indeed, for the ground state level of
Fig.~\ref{Profilesx01} ($x=0.1$, $L=6$), which is
marked with a circle in Fig.~\ref{enhancement}, the
giant vortex state has clearly been formed already.
On the contrary, for $x=0.5$, $L=5$, at $\Phi= 9 \,
\Phi_0$, marked with a square in
Fig.~\ref{enhancement}, superconductivity nucleates
in a uniform way ($\beta_A=1.01$) (see
Fig.~\ref{Profilesx05}).

The formation of the giant vortex state at high
$\Phi$ can be understood when writing the GL~free
energy~\cite{TinkBook,dGABook}
\begin{equation}
F=F_n+S \, <\alpha |\Psi|^2>+ S \, \frac{\hbar^2}{2
m^{\star}} \, <\left|
\left(- \imath \vec{\nabla}-
\frac{2 \pi}{\Phi_0} \vec{A} \right) \Psi
\right|^2> + \cdots \, ,
\label{GLfunctional}
\end{equation}
with the sum of the last two terms vanishing at the
phase boundary $T_{c}(H)$, where $\Psi
\rightarrow 0$. $F_n$ is the free energy of the
system in the normal state. The notation $<>$
denotes the average over the sample area $S$, and
$|\alpha|=\hbar^2/2 m^{\star}\xi^2(T)$. This yields
(see Eq.~(\ref{GL2})):
\begin{equation}
\frac{1}{\xi^2(T)}=\frac{<|\Psi|^2 \, \vec{v_s} \,^2>+
<\left(\vec{\nabla} |\Psi| \right)^2>}{<|\Psi|^2>}
\label{GLbalance}
\end{equation}
The solutions from Eq.~(\ref{psikummer}) which
fulfill the boundary conditions
(Eq.~(\ref{glbound})) in $r=r_i$ and $r=r_o$ need to
be inserted in this equation, and $v_s$ is
determined from Eq.~(\ref{fluxoid}). A fast
calculation shows that the relative contribution
from spatial gradients (the second term in the
numerator of Eq.~(\ref{GLbalance})) is 47~\% for the
lowest level of Fig.~\ref{Profilesx01}e ($x=0.1$,
$L=6$), while it is less than 4~\% for the ground
state of Fig.~\ref{Profilesx05}d ($x=0.5$, $L=5$).
The energetically most favourable balance between
these two contributions is strongly affected by the
boundary conditions (Eq.~\ref{glbound}). In
thin-wire loops, for example, bending of $|\Psi|$ on
a scale smaller than the coherence length $\xi$ will
result in a large contribution of
$<\left(\vec{\nabla} |\Psi|
\right)^2>$. Therefore, the $T_{c}$ reduction is
only determined by the averaged supercurrent kinetic
energy $<|\Psi|^2 \,
\vec{v_s}
\,^2>$ in this case, which is equivalent to the
London limit.

An examination of the supercurrent flow profiles
shows that, as in the case of a filled disk, there
is a paramagnetic contribution close to the sample
center for $L \neq 0$, while near the outer sample
interface diamagnetic currents are flowing. The
supercurrent density $J$ can be obtained by
inserting the general solution
(Eq.~(\ref{psikummer})) in the second GL~equation
(Eq.~(\ref{fluxoid})):
\begin{equation}
\vec{J} (L \, ,\Phi \,) = |\Psi|^2 \, \vec{v_s}
\propto
\left( L - \frac{\Phi}{\Phi_0} \right) \left(
\frac{\Phi}{\Phi_0} \right)^{L-\frac{1}{2}}
\exp \left( - \frac{\Phi}{\Phi_0} \right) \,
\left[ K(-N,L+1,\Phi / \Phi_0) \right]^2
\, \vec{e}_{\varphi}
\label{supercurrents}
\end{equation}
At a radius $r$, corresponding to integer flux
quanta $\Phi=L\,\Phi_0$ the current orientation
changes sign. It can happen, however, that the
switching radius $r<r_i$, lies inside the middle
opening. For the state $L=0$ there are only
diamagnetic currents. In contrast to this, for
example, for a loop with $x=0.5$, the states $L=2$
(Fig.~\ref{Profilesx05}a) and $L=3$
(Fig.~\ref{Profilesx05}b) do not carry paramagnetic
currents at $\Phi=9 \, \Phi_0$. At the ground state
level ($L=5$) (Fig.~\ref{Profilesx05}d), which is
very close to the London limit solution, $J \propto
v_s \propto (L-\Phi/\Phi_0)/r$
(Eq.~(\ref{fluxoid})). This is shown in the right
inset of Fig.~\ref{enhancement}, where the
supercurrent has been normalized to $-1$ at the
outer radius $r_o$. For the loop with $x=0.1$ (see
Fig.~\ref{Profilesx01}e), in the ground state $L=6$,
the dominating diamagnetic supercurrents are flowing
in the vicinity of $r=r_o$, the currents are
paramagnetic for slightly lower $r$, and $J$ is
vanishing in the core of the sample $r_i<r<r_o/3$
(left inset of Fig.~\ref{enhancement}).

As a last point, we discuss now the periodicity
$\Delta
\Phi$ of $T_{c}(\Phi)$, for the different $L$ states
in the loops. These results are shown in
Fig.~\ref{periods}. For the disk, the first cusp in
$T_{c}(\Phi)$, where $L$ goes from $0 \rightarrow
1$, occurs at $\Phi = 1.92 \, \Phi_0$, so $\Delta
\Phi (L=0)=3.85
\, \Phi_0$. The period $\Delta \Phi$ goes
down for increasing $L$, until it reaches the
asymptotic limit~\cite{Buisson90}:
\begin{equation}
\Delta \Phi = \Phi_0 \,
\left( 1+ (2 \, \eta \, \Phi / \Phi_0)^{-1/2}\right)
\label{asymperiod}
\end{equation}
Since $\eta$ is decreasing with $\Phi$, $\Delta
\Phi$ is a weakly decreasing function of $L$,
at high $\Phi$. The symbols in Fig.~\ref{periods}
correspond to the period $\Delta \Phi$ in units of
$\Phi_0$ for the different $L$ states. Since at the
highest $\Phi$, $T_{c}(\Phi)$ is calculated in steps
of $\approx 0.07 \, \Phi_0$, the error on $\Delta
\Phi$ is of this order. The filled squares are the
periods for the filled disk. The periodicity of
$T_{c}(\Phi)$ in the case of loops behaves
differently: for $x=0.1$ (filled up-triangles in
Fig.~\ref{periods}) $\Delta \Phi (L=1)=1.82 \,
\Phi_0$ is larger than for the filled disk, then $\Delta
\Phi (L=2)$ jumps below the corresponding value for the
disk, and for higher $L\geq 3$, the period $\Delta
\Phi$ matches with the disk behaviour. Consequently,
the giant vortex state builds up for $L \geq 3$,
i.e. $\Phi \gtrsim 5 \,
\Phi_0$. A similar analysis can be carried out for a loop
with $x=0.5$ (filled down-triangles in
Fig.~\ref{periods}). For low $L$, $\Delta \Phi
\approx 1.8 \, \Phi_0$, then $\Delta
\Phi(L)$ decreases substantially below the value for
the filled disk, before increasing again, until the
same period $\Delta \Phi(L)$ is reached for $L \geq
12$. The loop is in the giant vortex state when
$\Phi \gtrsim 17 \, \Phi_0$. For loops made of even
thinner wires ($x=0.7$ and $x=0.9$) (open symbols in
Fig.~\ref{periods}), $T_{c}(\Phi)$ stays periodic up
to $L=13$. The constant period $\Delta
\Phi$ corresponds to
$\Delta \Phi_m=\left((1+x)/2 \right)^2 \, \Delta
\Phi= 1 \, \Phi_0$, as it should be in the London
limit for $x \rightarrow 1$, according to
Eq.~(\ref{groffexp}).

In summary, we have solved the linearized
GL~equation for loops of different wire width, with
Neumann boundary conditions at both the outer and
the inner loop radius. The critical fields
$H_{c3}^{*}(T)$ are always above the $H_{c3}(T)=1.69
\, H_{c2}(T)$, the surface critical field for a
semi-infinite superconducting slab in contact with
vacuum. The ratio $H_{c3}^{*}(T)/H_{c2}(T)$
increases when the size of the middle opening grows,
i.e. in a sample topology with a large
surface-to-volume ratio the nucleation field is
strongly enhanced. Depending on the ratio inner to
outer radius $r_i/r_o$ of the loops, and on the
applied magnetic flux, $T_{c}(\Phi)$ shows a
different behaviour: in thin-wire loops, the
background of $T_{c}$ is parabolic (characteristic
for 2D behaviour) and the Little-Parks $T_{c}(\Phi)$
oscillations are perfectly periodic. This behaviour
can be described in the London limit. For loops with
only a very small opening, the period of the
$T_{c}(\Phi)$ oscillations is decreasing with $\Phi$
and the background $T_{c}$ reduction is quasi-linear
(3D regime, e.g., like for a disk). Intermediate
loops (for instance $r_i/r_o=0.5$) show a 2D-3D
cross-over between the two regimes at a certain
applied flux $\Phi$ (corresponding to
$w=r_o-r_i=(1-x)\, r_o
\approx 2 \, \xi(T)$), similar to the dimensional
transition in thin films subjected to a parallel
field. As soon the 3D regime is reached, a giant
vortex state is created, where only a sheath close
to the sample's outer interface is superconducting.
The opening in the middle of the loop does not play
a role anymore: $T_{c}(\Phi)$ for the loop and for
the disk become identical.

\section*{Acknowledgments}
The authors wish to thank H.~J.~Fink, T.~Puig,
J.~G.~Rodrigo, J.~T.~Devreese, V.~M.~Fomin,
K.~Temst, and J.~O.~Indekeu for stimulating
discussions. This work has been supported by the
Belgian IUAP, the Flemish GOA and FWO-programmes,
and by the ESF programme VORTEX.
\\

\noindent $^{*}$ e-mail: Vital.Bruyndoncx@fys.kuleuven.ac.be

\begin{figure}
\caption{Calculated energy level scheme (dashed lines)
for a superconducting loop with different ratio of
inner to outer radius $x=r_i/r_o$: a)~$x=0.1$,
b)~$x=0.3$, c)~$x=0.5$, d)~$x=0.7$, and e)~$x=0.9$.
The lowest level for each magnetic flux
$\Phi/\Phi_0$ corresponds to the highest possible
temperature $T_{c}(H)$ for which superconductivity
can exist. A state with phase winding number $L=0$
is formed at $T_{c}(\Phi
\approx 0)$, and at each cusp in $T_{c}(H)$ the
system makes a transition $L\rightarrow L+1$,
indicating the entrance of an extra vortex. The
solid and dotted lines correspond to $H_{c2}(T)$ and
$H_{c3}(T)$, respectively.}
\label{elevels}
\end{figure}

\begin{figure}
\caption{Order parameter distribution $|\Psi|$ for
$x=r_i/r_o=0.1$, at a fixed $\Phi=9 \, \Phi_0$.
a)~$L=2$, b)~$L=3$, c)~$L=4$, d)~$L=5$, and
e)~$L=6$. The latter ($L=6$) corresponds to the
ground state level $|-\alpha_{LLL}(9
\,\Phi_0)|$, where the sample is in the giant vortex
state.}
\label{Profilesx01}
\end{figure}

\begin{figure}
\caption{Order parameter distribution $|\Psi|$ for $x=0.5$,
at a fixed $\Phi=9 \, \Phi_0$. a)~$L=2$, b)~$L=3$,
c)~$L=4$, d)~$L=5$, and e)~$L=6$. The state with
$L=5$ is only slightly modulated ($|\Psi| \approx$
constant: $\beta_A=1.01$) and corresponds to the
ground state level $|-\alpha_{LLL}(9 \,\Phi_0)|$. }
\label{Profilesx05}
\end{figure}

\begin{figure}
\caption{Inverse enhancement factor
$\eta^{-1}=\epsilon(H_{c3}^{*})/\epsilon(H_{c2})$
for loops with different aspect ratio, compared to
the case of a disk and an antidot. The horizontal
dashed line at $\eta^{-1}=0.59=1/1.69$ corresponds
to $H_{c3}(T)/H_{c2}(T)=1.69$ for a plane
superconductor/vacuum boundary. The insets show the
supercurrent $J$ profiles for the ground states at
$\Phi= 9 \, \Phi_0$ for $x=r_i/r_o=0.1$ (left) and
$x=0.5$ (right). These are obtained from
Eq.~(\ref{supercurrents}) and are normalized to $-1$
at the outer radius $r=r_o$. The corresponding
$|\Psi|$ profiles are plotted in
Fig.~\ref{Profilesx01}e, and
Fig.~\ref{Profilesx05}d, respectively.}
\label{enhancement}
\end{figure}

\begin{figure}
\caption{The period $\Delta \Phi$ of the phase boundary
$T_{c}(\Phi)$ in units of the flux quantum $\Phi_0$
as a function of the phase winding number $L$. The
data, for several of the loops is shown as a symbol,
and is compared to the period in a filled disk
(filled square). The interconnecting lines are only
a guide to the eye.}
\label{periods}
\end{figure}

\end{document}